\documentclass[preprint]{aastex62}

\newcommand{\speed}[1]{#1 km~s${}^{-1}$}

\newcommand{\nfig}[1]{Figure~\ref{#1}}

\usepackage{amsmath}

\shorttitle{An Erupting Mini-filament Driven Two-Sided-Loop Jet}
\shortauthors{Shen et al.}

\begin{document}

\title{Stereoscopic Observations of an Erupting Mini-filament Driven Two-Sided-Loop Jet and the Applications for Diagnosing Filament Magnetic field}

\correspondingauthor{Yuandeng Shen}
\email{ydshen@ynao.ac.cn}

\author[0000-0001-9493-4418]{Yuandeng Shen}
\affiliation{Yunnan Observatories, Chinese Academy of Sciences,  Kunming, 650216, China}
\affiliation{Center for Astronomical Mega-Science, Chinese Academy of Sciences, Beijing, 100012, China}
\author{Zhining Qu}
\affiliation{School of Automation and Information Engineering, Sichuan University of Science \& Engineering, Zigong 643000, China}
\author{Ding Yuan}
\affiliation{Institute of Space Science and Applied Technology, Harbin Institute of Technology, Shenzhen, 518055, China}
\author{Huadong Chen}
\affiliation{University of Chinese Academy of Sciences, Beijing, China}
\affiliation{Key Laboratory of Solar Activity, National Astronomical Observatories, Chinese Academy of Sciences, Beijing 100012, China}
\author{Yadan Duan}
\affiliation{Yunnan Normal University, Department of Physics, Kunming 650500, China}
\author{Chengrui Zhou}
\affiliation{Yunnan Observatories, Chinese Academy of Sciences,  Kunming, 650216, China}
\affiliation{University of Chinese Academy of Sciences, Beijing, China}
\author{Zehao Tang}
\affiliation{Yunnan Observatories, Chinese Academy of Sciences,  Kunming, 650216, China}
\affiliation{University of Chinese Academy of Sciences, Beijing, China}
\author{Jin Huang}
\affiliation{Yunnan Observatories, Chinese Academy of Sciences,  Kunming, 650216, China}
\affiliation{University of Chinese Academy of Sciences, Beijing, China}
\author{Yu Liu}
\affiliation{Yunnan Observatories, Chinese Academy of Sciences,  Kunming, 650216, China}

\begin{abstract}
The ubiquitous solar jets or jet-like activities are generally regarded as an important source of energy and mass input to the upper solar atmosphere and the solar wind. However, questions about their triggering and driving mechanisms are not completely understood. By taking advantage of high temporal and high spatial resolution stereoscopic observations taken by the {\em Solar Dynamic Observatory} ({\em SDO}) and the {\em Solar Terrestrial Relations Observatory} ({\em STEREO}), we report an intriguing two-sided-loop jet occurred on 2013 June 02, which was dynamically associated with the eruption of a mini-filament below an overlying large filament, and two distinct reconnection processes are identified during the formation stage. The {\em SDO} observations reveals that the two-sided-loop jet showed a concave shape with a projection speed of about \speed{80 -- 136}. From the other view angle, the {\em STEREO} ahead observations clearly showed that the trajectory of the two arms of the two-sided-loop were along the cavity magnetic field lines hosting the large filament. Contrary to the well-accepted theoretical model, the present observation sheds new light on our understanding of the formation mechanism of two-sided-loop jets. Moreover, the eruption of the two-sided-loop jet not only supplied mass to the overlying large filament, but also provided a rare opportunity to diagnose the magnetic structure of the overlying large filament via the method of three-dimensional reconstruction. 
\end{abstract}

\keywords{Sun: activity --- Sun: flares --- Sun: filaments, prominences --- Sun: magnetic fields --- Sun: coronal mass ejections (CMEs)} 

\section{Introduction}
Solar jets are collimated, beam-like plasma outflows along magnetic field lines \citep{1992PASJ...44L.173S,1994ApJ...431L..51S}; they are omnipresent in the solar atmosphere and can be observed from the photosphere to several solar radii in the outer corona \citep{1973SoPh...28...95R,1996ApJ...464.1016C,1999ApJ...513L..75C,1998ApJ...508..899W,2009ApJ...707L..37L,2016ApJ...823..129A,2018ApJ...854...92T}, and even at one astronomical unit (AU) near the Earth \citep{2006ApJ...639..495W,2008ApJ...675L.125N,2012ApJ...750...50N}. Previous observations indicate that solar jets are commonly associated with magnetic flux emergence, cancellation, and micro-flares at their eruption source region \citep{2007A&A...469..331J,2017ApJ...844..131P,2018ApJ...864...68S}. Statistical studies show that solar jets have a speed of about \speed{200}, widths in the range of 2000-200000 km, and lifetimes from minutes to tens of minutes \citep{2009SoPh..259...87N,2011ApJ...735L..43S,2013ApJ...776...16P,2016SSRv..201....1R}. Studies in the past few decades have indicated that jet phenomenon is very important for our understanding of solar physics. For example, the high occurrence frequency and high speed of solar jets imply that they may important for accounting the enigmatic problems of coronal heating and the acceleration of the fast solar wind \citep{2007Sci...318.1591S,2014Sci...346A.315T}. The eruption of small-scale solar jets can not only lead to the formation of large-scale coronal magnetic reconfiguration phenomena such as filament eruptions and coronal mass ejections \citep[e.g.,][]{1998ApJ...508..899W,2008ApJ...677..699J,2012ApJ...745..164S,2018NewA...65....7T,2018JPhCS1100a2024S,2019arXiv190707310D}, but also can excite large-scale coronal waves directly and indirectly \citep{2018ApJ...861..105S,2018MNRAS.480L..63S,2018ApJ...860L...8S,2019ApJ...873...22S}. In addition, investigating solar jets can open doors to understand highly complex astrophysical jets that are commonly associated with compact central objects such as black holes, neutron stars or pulsars \citep{2014ARA&A..52..529Y}. 

Early low resolution soft X-ray observations revealed two types of solar jets based on their morphology, namely anenome jets and two-sided loop jets \citep{1994xspy.conf...29S}. An anenome jet shows as an outward collimated plasma beam and is accompanied by a bright point at the edge of the jet's base, while a two-sided loop jet shows as a pair of plasma beams at both sides of the eruption source region. In the classical jet model \citep{1995Natur.375...42Y}, the two types of solar jets are all interpreted through magnetic reconnection (that is (a physical process that breaks and reconnects of oppositely directed magnetic field lines in highly conducting plasmas, in which the magnetic energy is converts to plasma kinetic and internal energy.) According to the model, an anenome (two-sided loop) jet is produced when an emerging loop reconnects with the ambient pre-existing open (horizontal) field lines, in which the acceleration of the jet plasma is due to the tension force of the newly formed field lines in the reconnection (i.e., slingshot effect). In some collimated anenome jets, both hot and cool plasma beams can be observed, and the appearance of the cool component is delayed with respect to the hot one a few minutes \citep{1994ApJ...425..326S,1999SoPh..190..167A,2007A&A...469..331J}. The hot plasma component is simply explained as the heated coronal plasma flow due to the magnetic reconnection, but the generation of the cool plasma flow has different interpretations such as the cooling of the earlier hot component \citep{1994ApJ...425..326S,1999SoPh..190..167A,2007A&A...469..331J}, different Alfv\'{e}n velocities in cool and hot plasmas \citep{2008ApJ...683L..83N}, and chromospheric cool plasmas carried up with expanding loops \citep{1995Natur.375...42Y}. So far, observations on cool component plasma flow in two-sided loop jets are still very scarce, although numerical simulation had predicted its existence \citep{1995Natur.375...42Y}.

Recent high temporal and high spatial resolution observations indicated that solar jets are more complicated than it used to be thought. New observations based on high resolution data show that the eruption characteristics of most solar jets are largely different to the physical picture of anenome jets, and those anemone jets with a broad spire produced by an eruption at their base were named blowout jets \citep{2010ApJ...720..757M}. Many observations have indicated that the eruption of blowout jets are tightly associated with the eruption of mini-filaments (MF) in the source region \citep{2012ApJ...745..164S,2017ApJ...851...67S,2014ApJ...783...11A,2015Natur.523..437S,2014ApJ...796...73H,2017ApJ...835...35H,2015ApJ...814L..13L,2017ApJ...842L..20L,2017ApJ...834...79Z,2019ApJ...872...87L,2019arXiv190707310D}. Besides the external reconnection as described in the anenome jet model, the strongly sheared core field of blowout jets still undergoes an internal reconnection process, and the appearance of the cool component has been shown in many cases to be the eruptingÊfilament material from the eruption source region rather than the classical Êmechanism introduced earlier \citep[e.g.,][]{2012ApJ...745..164S,2017ApJ...851...67S}.

So far, the classical model of two-sided loop jet has been supported by many observations \citep{1998SoPh..178..173K,2013ApJ...775..132J,2018ApJ...853L..26H,2018ApJ...861..108Z}; however, recent high resolution observations reveal that two-sided loop jets are highly associated with the eruption of mini-filaments. For example, \cite{2017ApJ...845...94T} reported a two-sided loop jet that was produced by the magnetic reconnection between two adjacent filamentary chromospheric structures; \cite{2019ApJ...871..220S} found two-sided-loop jet that was driven by the eruption of a mini-filament, in which the erupting filament interacts with an overlying horizontal coronal loop at about 30 Mm in altitude and producing the two-sided-loop jet. To advance our understanding of two-sided loop jets, in this paper, we report the detailed formation process of an intriguing two-sided loop jet, by taking advantage of the high resolution stereoscopic observations taken from different view angles. It is evidenced that the birth of the two-sided loop jet not only underwent two reconnection processes, but also involved the eruption of a mini-filament in the eruption source region. These findings sheds new light on our understanding of the physical nature of two-sided loop solar jets.

\section{Results}
The two-sided loop jet occurred on 2013 June 02 below a large filament (LF) located in the quiet-Sun region, which was observed simultaneously by space satellites including the {\sl Solar Dynamic Observatory} \citep[{\sl SDO};][]{2012SoPh..275....3P} and the {\sl Solar TErrestrial RElations Observatory} ahead \citep[{\sl STEREO}-A;][]{2008SSRv..136....5K} from two different observation angles. The extreme ultraviolet (EUV) observations of 94 \AA\, 171 \AA\, 193 \AA\, 211 \AA\, 304 \AA\, and 335 \AA\ taken by the {\sl SDO}'s Atmospheric Imaging Assembly \citep[AIA;][]{2012SoPh..275...17L} are used, and their strong responses to logarithmic temperatures (in Kelvin) are of about 6.8, 5.8, 6.2, 6.3, 4.7, and 6.4 respectively. The cadence and pixel size of the AIA images are respectively 12 seconds and 441 kilometers. The line-of-sight (LOS) magnetograms taken by the {\sl SDO}'s Helioseismic and Magnetic Imager are also used, whose cadence and pixel size are respectively 45 seconds and 368 kilometers. The {\sl STEREO}-A provided EUV 195 \AA\ and 304 \AA\ images with a pixel size of 1107 kilometers, and their cadences are of 5 and 10 minutes, respectively. In addition, we also used the H$\alpha$ images from the Global Oscillation Network Group \citep[GONG;][]{1996Sci...272.1284H}.

\begin{figure}
\epsscale{1}
\figurenum{1}
\plotone{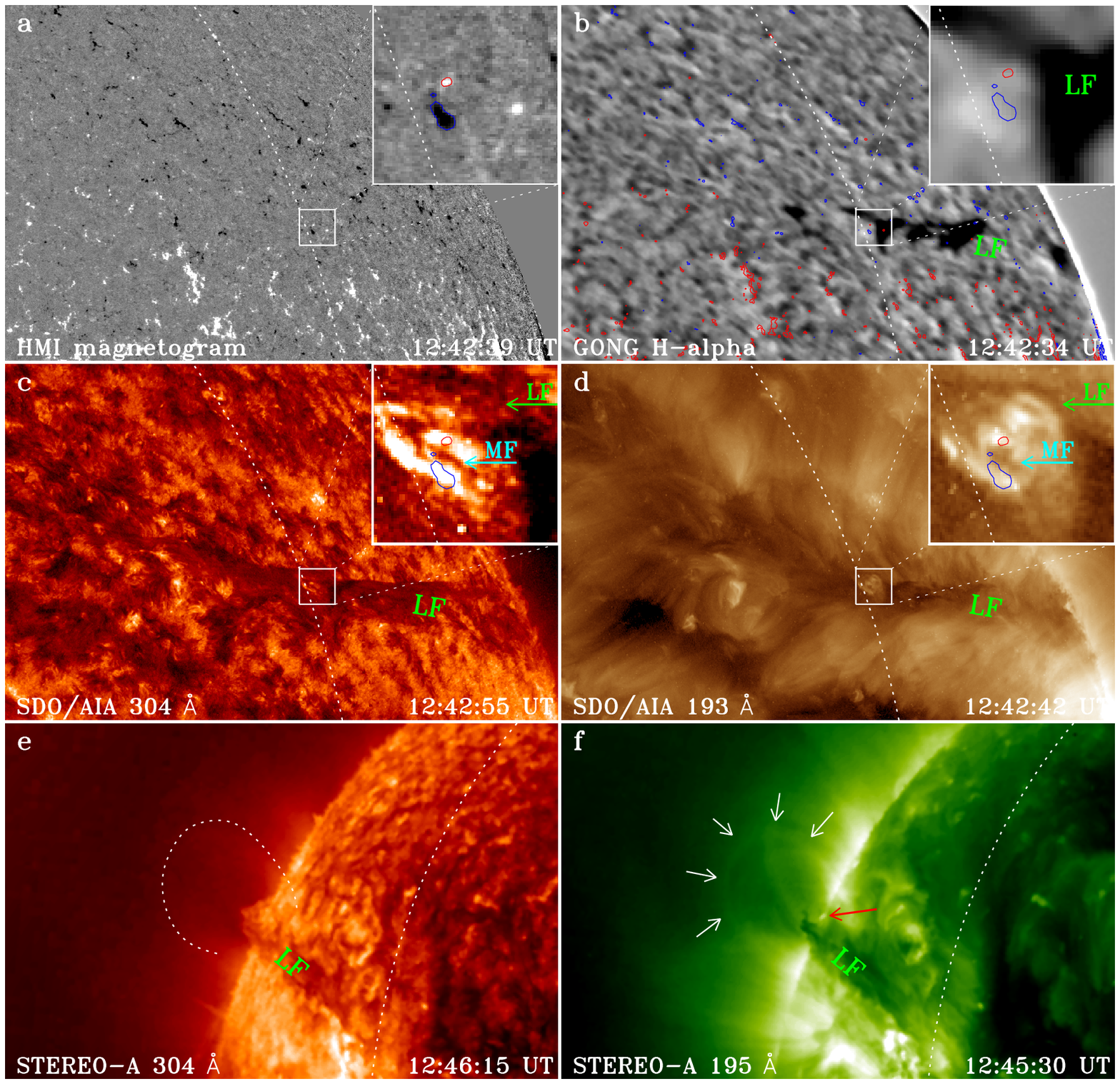}
\caption{Pre-eruption magnetic configuration. (a) HMI LOS magnetogram in which bright (black) patches are positive (negative) polarities. (b) GONG H$\alpha$ image shows LF, in which the red (blue) contours indicate the magnetogram in (a) at 100 (-100) Gauss. (c--d) AIA 304 \AA\ and 193 \AA\ images shows the coronal condition. (e--f) {\em STEREO}-A 304 \AA\ and 195 \AA\ images. The close-up shot of the eruption source region is plotted at the top-right in each panel from (a) to (d), and the red (blue) contour shows the small positive (negative) polarity at 50 (-50) Gauss. The cyan and green arrows in the (c) and (d) indicate the MF and the LF, respectively. The dotted curve in (e) indicates the cavity profile determined from (f), which is also indicated by the white arrows in (f). The red arrow in (f) points to the eruption source region. The white dotted curves in the AIA images marks the east disk limb in the FOV of the {\em STEREO}-A, while that in the {\em STEREO}-A images marks the west disk limb in the FOV of the {\em SDO}. The field of view (FOV) of a--d is 620\arcsec $\times$ 400\arcsec, while that for (e--f) is 930\arcsec $\times$ 600\arcsec  (1 arcsec corresponds to about 735 kilometers on the Sun).
\label{fig1}}
\end{figure}

The eruption source region of the two-sided loop jet was a small bipolar region with a MF resided in the magnetic neutral region (see \nfig{fig1} a--d and the inset in each panel), above which was the long LF that was composed of several dark segments in the GONG H$\alpha$ image (see \nfig{fig1}b). The MF can well be identified in the AIA 304 \AA\ and 193 \AA\ images, and it is indicated by the cyan arrows in \nfig{fig1}c and d. On 2013 June 02, the separation angle between {\sl SDO} and {\sl STEREO}-A was about 140 degrees. The eruption source region of the two-sided loop jet located coincidentally on the east disk limb in the field-of-view (FOV) of {\sl STEREO}-A. It is interesting that the main axis of the LF was almost parallel to the line-of-sight (LOS), and the cavity that hosted the LF at its bottom can well be identified as a circular loop structure over the east disk limb in the {\sl STEREO}-A 195 \AA\ image (see \nfig{fig1} e and f).

\begin{figure}
\epsscale{1}
\figurenum{2}
\plotone{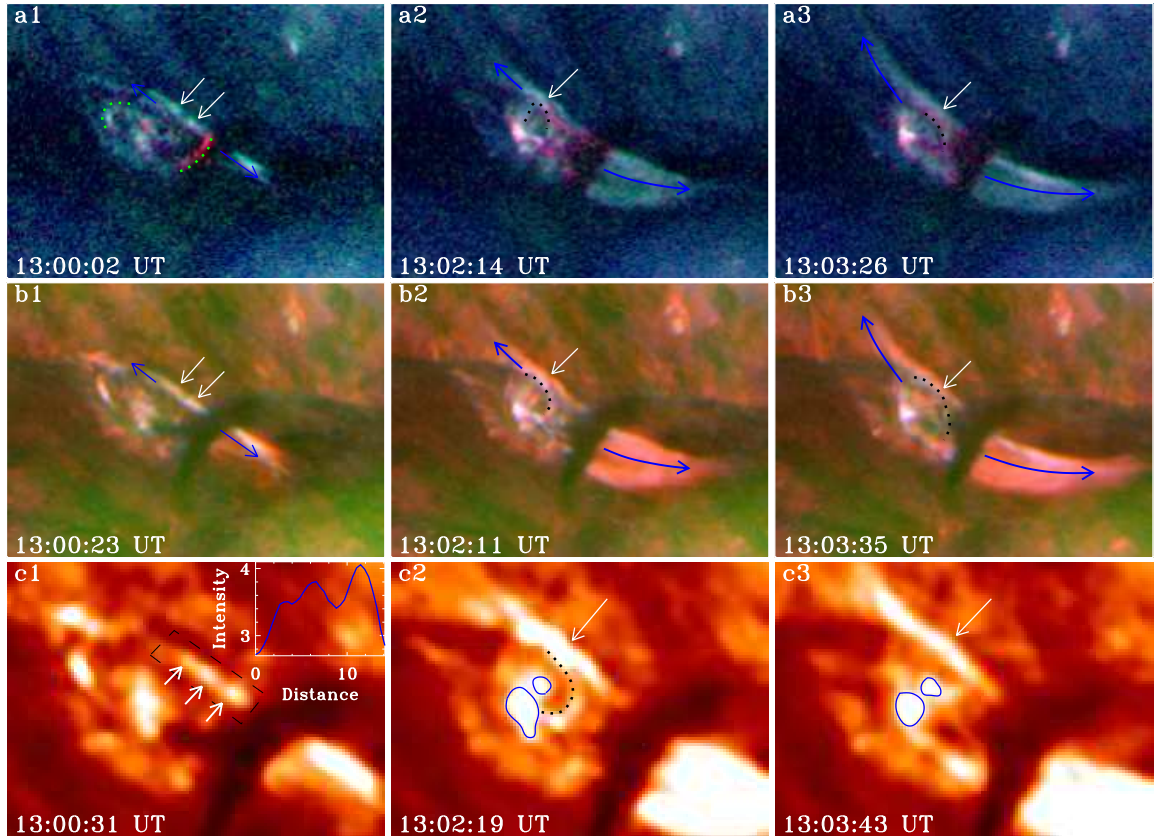}
\caption{Eruption details of the two-sided-loop jet in AIA images. (a1--a3) composite high temperature images made from the AIA 94 \AA\ (red), 335 \AA\ (green), and 193 \AA\ (blue) channels. (b1--b3) composite low temperature images made from the AIA 304 \AA\ (red), 211 \AA\ (green), and 171 \AA\ (blue) channels. (c1--c3) AIA 304 \AA\ images. The white arrows in this figure indicate the reconnection current sheet, while the paired blue arrows indicate the two arms of the two-sided-loop jet. The green dotted curves in (a1) mark the two-sided-expanding loop structures, while the black dotted curves in a2--a3, b2--b3, and c2 mark the rising MF. The three white arrows in (c1) indicate the magnetic islands along the current sheet, and its relative intensity profile is plotted as an inset on the top-right corner. The black dashed box shows the region used for obtaining the AIA lightcurve in \nfig{fig3} (c). The two bright flare ribbons are outlined by the blue contours in (c2) and (c3). The FOV of the top and middle rows is 132\arcsec $\times$ 96\arcsec, while that for the bottom row is 79\arcsec $\times$ 56\arcsec. An animation made from the composite high and low temperature images is available and runs from ~12:40 to ~13:40 UT.
\label{fig2}}
\end{figure}

The formation of the two-sided loop jet can be divided into two stages, and we show these processes with composite high and low temperature images in the top and middle rows of \nfig{fig2}, respectively. To make a composite image, three simultaneous AIA images of different wavelengths are put into the red, green, and blue channels of a true color image. Here, the strong response to logarithmic temperature of the composite high and low temperature images are in the ranges of 6.2--6.8 and 4.7--5.8, respectively. The first stage was about from 12:59:00 UT to 13:01:30 UT. During this stage, several observing features appeared, including 1) brightening in the source region; 2) a pair of hot loop structures moved in opposite directions (see the dotted green curves in \nfig{fig2} a1); 3) a linear bright structure between the rising loop and the LF (indicated by the white arrows in \nfig{fig2} a1 and b1); and 4) a pair of small plasma flows as indicated by the two blue arrows \nfig{fig2} a1 and b1. These characteristics have some aspects in common with the physical picture described in the numerical model of two-sided loop jets \citep{1995Natur.375...42Y}, where an emerging closed loop reconnects with the overlying horizontal magnetic fields. Here, the linear bright structure can be regarded as the current sheet in which the rising loop reconnected with the overlying magnetic field of the LF, and the ejecting plasma flows in opposite directions are the heated plasma outflows accelerated by the tension force of the newly formed magnetic field lines.
 
The second stage started at about 13:01:30 UT from the rising of a bright loop-like structure which then interacted with the overlying LF. Similar to the first stage, besides the brightening in the eruption source region, a current sheet also formed between the rising loop structure and the overlying magnetic field of the LF (indicated by the white arrows in \nfig{fig2} a2, a3, b2, and b3), where magnetic reconnection occurred and launched the intriguing two-sided loop jet (indicated by the paired blue arrows in \nfig{fig2} a2, a3, b2, and b3). The most striking feature observed in the second stage is the appearance of dark cool filament material at the lower edge of the bright loop structure, which erupted along with the loop structure and then merged into the two arms of the jet (see the dotted black curve in \nfig{fig2} a2, a3, b2, and b3). The close-up shots of the jet's eruption source region are shown in the bottom row of \nfig{fig2} with AIA 304 \AA\ images. It is clear that the current sheets in the two eruption stages are all composed of multiple magnetic islands whose projection widths are in the range of 1300 to 3000 kilometers. This is consistent with the numerical results  \citep{1995Natur.375...42Y,2017ApJ...841...27N} and observations \citep{2017ApJ...851...67S,2019ApJ...870..113Z}, confirming the occurrence of reconnection processes in the two eruption stages. Another feature identified in the AIA 304 \AA\ images is the appearance of two bright points during the second eruption stage, which located on both sides of the pre-eruption MF. The two bright points can be regarded as the conjugated flare double ribbons as observed in many large-scale filament eruptions. Therefore, the appearance of the two bright points and the dark cool material in the second stage together suggest the eruption of the MF. More eruption details can be found by seeing the animation (animation1.mpg) in the online journal.

\begin{figure}
\epsscale{1}
\figurenum{3}
\plotone{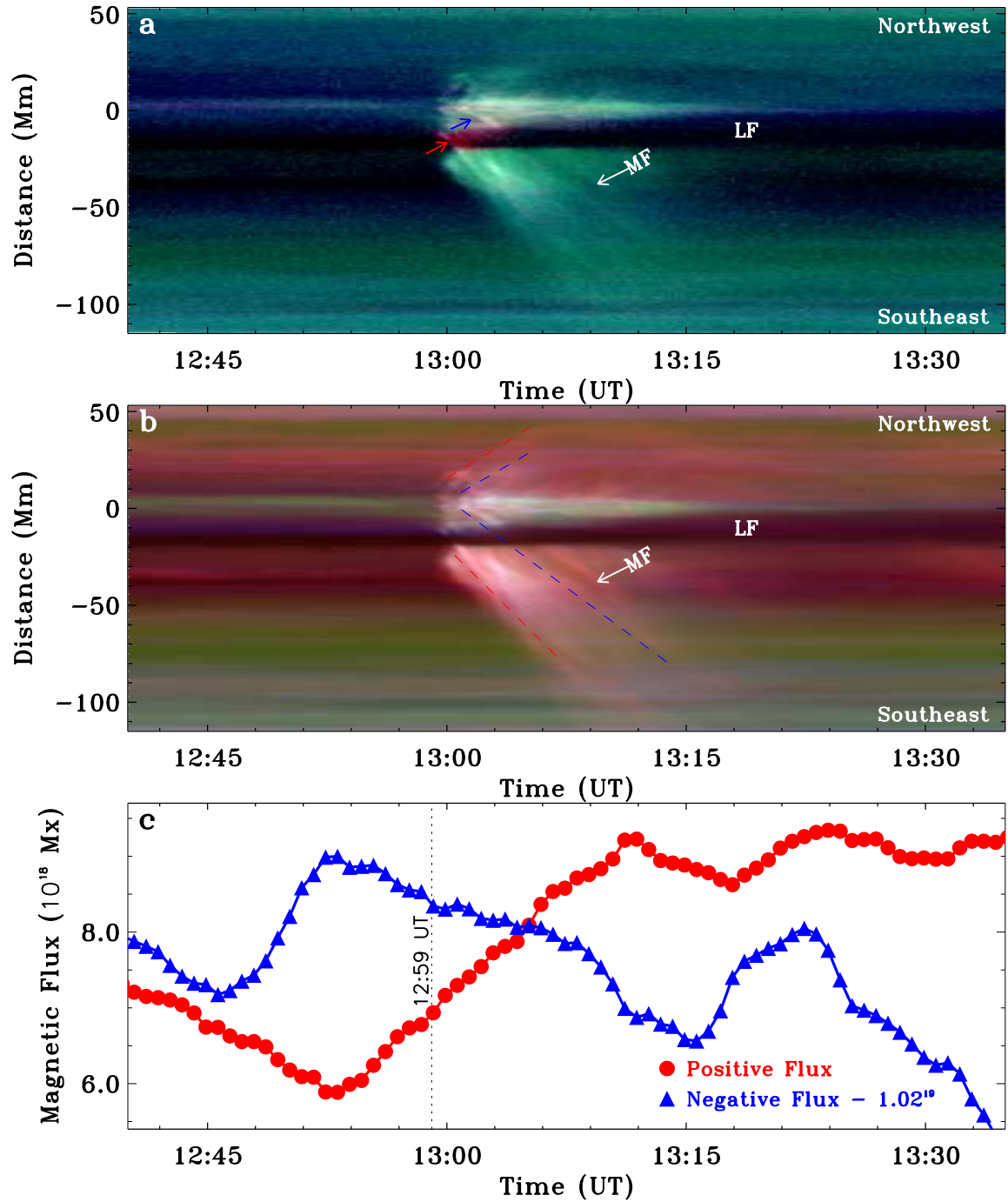}
\caption{TDS plots along the jet's main axis and the magnetic flux variations in the eruption source region. (a) composite TDS plot made from AIA high temperature channels (94 \AA\ (red), 335 \AA\ (green), and 193 \AA\ (blue)). (b) composite TDS plot made from AIA low temperature channels (304 \AA\ (red), 211 \AA\ (green), and 171 \AA\ (blue)). The red and blue arrows in (a) indicate the first and the second reconnection processes, and the corresponding ejecting plasmas in opposite directions are indicated by the red and blue dashed lines in (b), respectively. The white arrows in (a) and (b) indicate the erupting cool plasma of the MF. (c) a plot of the magnetic flux variation in the small box region as shown in \nfig{fig1} (a). Red circle and blue triangle show the positive flux and the absolute value of the negative flux, respectively. The vertical black dotted line indicates the start time of the two-sided-loop jet.
\label{fig3}}
\end{figure}

The eruption details can also be identified in the time-distance stack (TDS) plots made from the time-sequence composite AIA images along the center axis of the two-sided loop jet. To obtain a TDS plot, one need to first extract the one-dimensional intensity profile along a specific path, then a two-dimensional TDS can be obtained by stacking all the one-dimensional intensity profiles in time. The top and middle rows of \nfig{fig3} are the composite high and low temperature TDS plots respectively.  In the TDS plots, the LF shows as a thick and dark horizontal bar, and the two reconnection (or ejection) processes are respectively indicated by the red and blue arrows in \nfig{fig3} a. Especially, the erupting dark filament material can also be identified clearly during the second reconnection process (see the white arrows in \nfig{fig3} a and b). By fitting the outward erupting plasma flows of the jet's arms with a linear function, it is obtained that the projection speeds in the plane of the sky of the north (south) arm formed in the first stage was about \speed{84 (136)}, while that formed in the second stage was about \speed{80 (102)}.

The time evolution of the magnetic fluxes within the eruption source region are plotted in \nfig{fig3}c to investigate the triggering mechanism of the two-sided loop jet. From about 12:46:00 UT before the eruption, the positive flux decreased quickly and reached its nadir at about 12:52:00 UT, then it started a long increasing phase. In the meantime, the variation trend of the negative flux was opposite to the positive flux. It first increased quickly and reached its peak at 12:52:00 UT; then it started a long decreasing phase. Such changing pattern of magnetic fluxes suggests the flux emergence and cancellation between the positive and negative fluxes. Here, the negative flux first emerged and cancelled with the surrounding positive flux, then after 12:52:00 UT the positive flux began to emerge and cancelled with the negative flux. It is noted that the start time of the flux cancellation was before the onset time of the two-sided loop jet a few minutes, which may indicate the rising time of the loop before its interaction with the overlying magnetic field of the LF. The close temporal and spatial relationships between the magnetic flux and the two-sided loop jet suggests that the present two-sided loop jet was probably triggered by the alternative flux emergences and cancellations.

\begin{figure}
\epsscale{1}
\figurenum{4}
\plotone{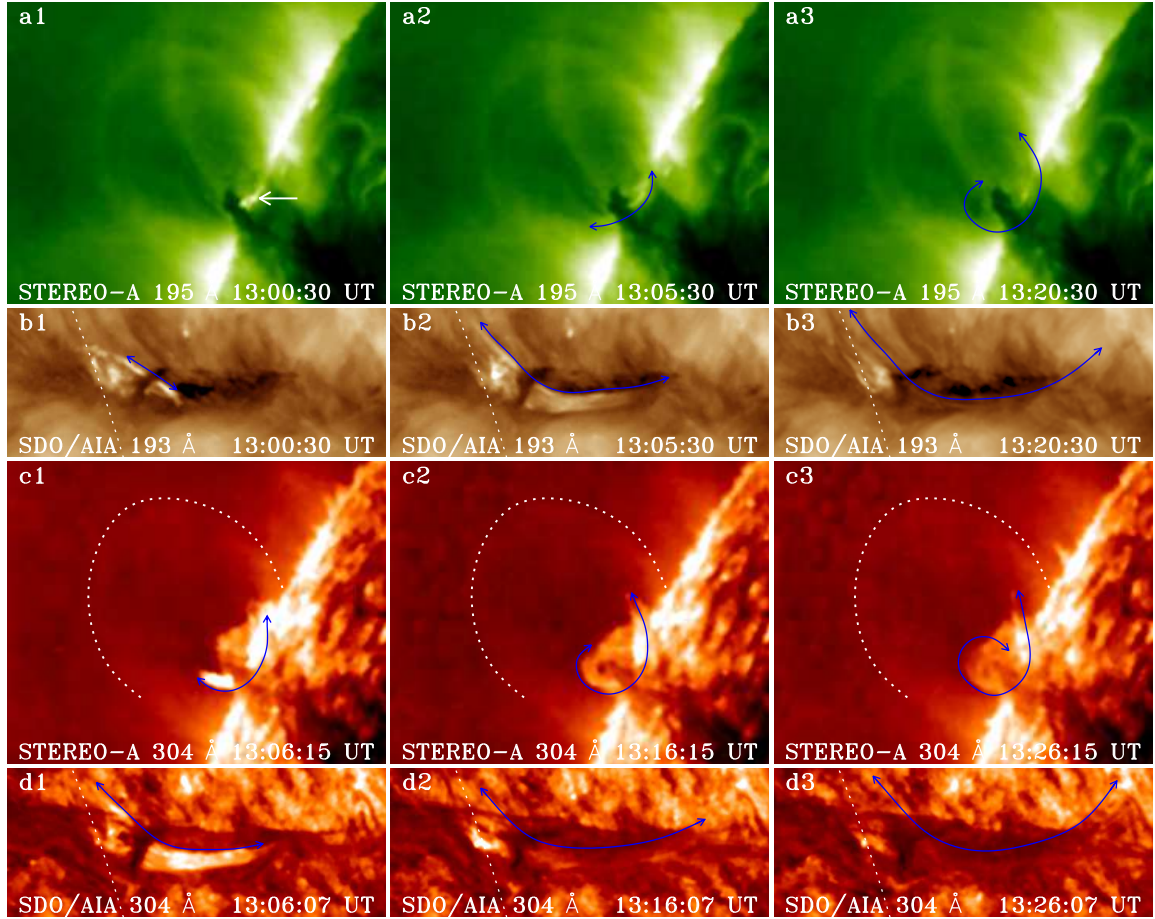}
\caption{Eruption details of the two-sided-loop jet in {\em STEREO}-A and {\sl SDO}/AIA images. (a1--a3) and (c1--c3) are {\em STEREO}-A 195 \AA\ and 304 \AA\ images, while (b1--b3) and (d1--d3) are AIA 193 \AA\ and 304 \AA\ images, respectively. The white arrow in (a1) indicates the beginning of the two-sided-loop jet, whose ejection trajectory and direction are indicated by the blue curved arrows in other panels. The white dotted curves in (c1--c3) are the cavity profile determined from (a1). The white dotted curves in the AIA images marks the east disk limb in the FOV of {\em STEREO}-A. The FOVs for the {\em STEREO}-A and AIA images are 390\arcsec $\times$ 320\arcsec and 280\arcsec $\times$ 112\arcsec, respectively. An animation made from the {\em STEREO}-A 195 \AA\ and 304 \AA\ time sequence images is available and runs from ~12:40 to ~14:45 UT.
\label{fig4}}
\end{figure}

The eruption of the two-sided loop jet was also observed by the {\sl STEREO}-A from the other observation angle, and its morphologies at different times are compared with the {\sl SDO}/AIA observations in \nfig{fig4}. In the FOV of {\sl STEREO}-A, it can be seen that the eruption source region just located at the east limb as indicated by the white arrow in \nfig{fig4} a1. The trajectory of the two-sided loop jet showed as a spiral structure almost perpendicular to the axis of the large filament (see the blue curves in the top and the third rows of \nfig{fig4} and the animation (animation2.mpg) available in the online journal). However, in the {\sl SDO}/AIA images, the two-sided loop jet exhibited as a concave structure and with a small acute angle with respect to the main axis of the LF (see the blue curves in the second and the bottom rows of \nfig{fig4}). These results indicate that the two-sided loop jet occurred under the large filament and with its two arms along the magnetic field lines of the cavity that hosted the material of the LF at its bottom. Therefore, the two-sided loop jet not only traced out the magnetic structure of the LF, but also injected mass into the LF. 

\begin{figure}
\epsscale{1}
\figurenum{5}
\plotone{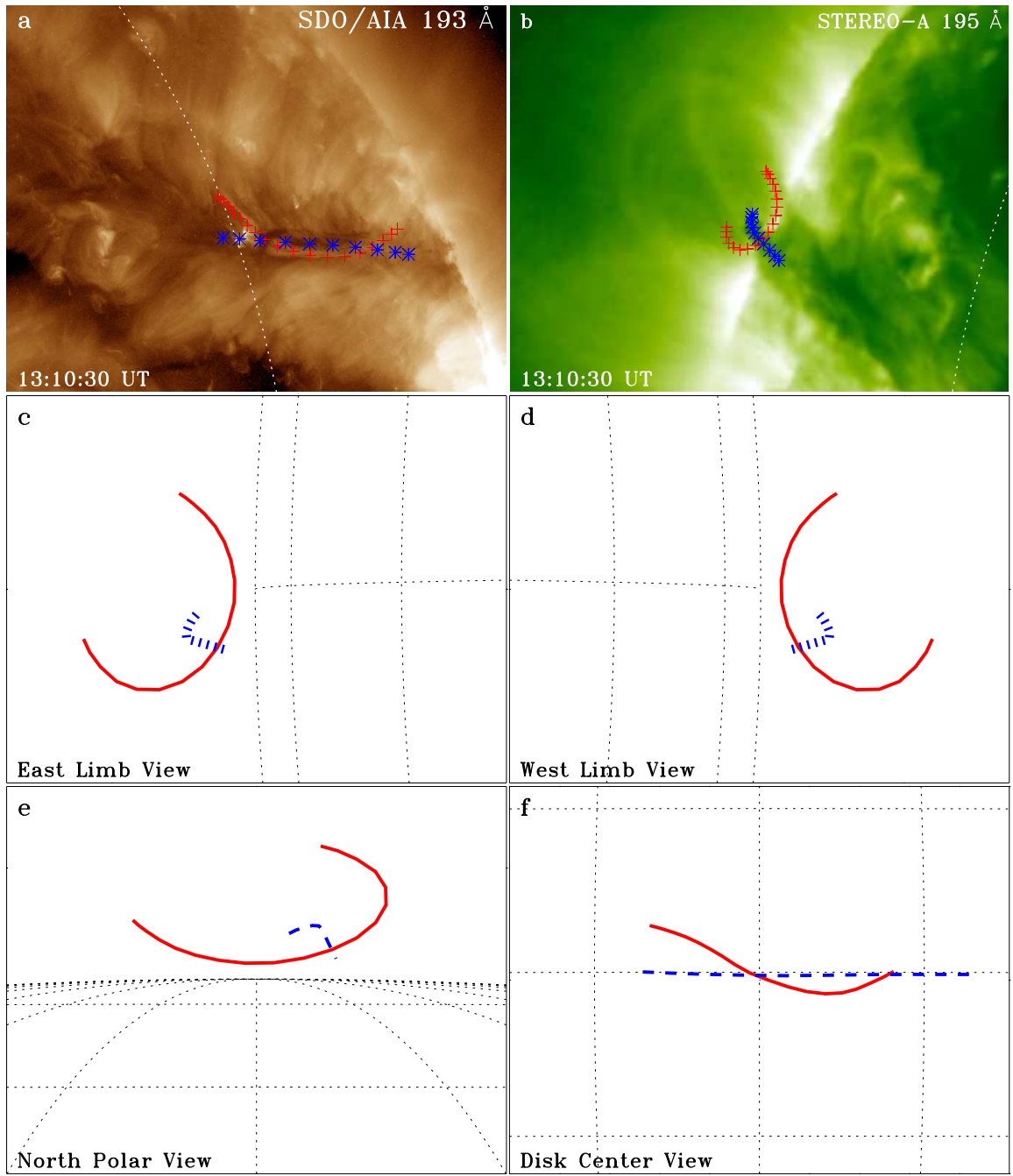}
\caption{Three-dimensional reconstruction of the trajectories of the two-sided-loop jet and the main axis of the LF. Top row shows the paired simultaneous SDO/AIA 193 \AA\ and {\em STEREO}-A 195 \AA\ images at 13:10:30 UT. The middle row shows the projected trajectories of the two-sided-loop jet (red) and the main axis of the LF at the east limb (panel c) and west limb (panel d) of the solar disk, while the bottom row shows them at the north polar (panel e) and the disk center (panel f).
\label{fig5}}
\end{figure}

By using the paired {\sl SDO} and {\sl STEREO}-A images at 13:10:30 UT, the true three-dimensional trajectories of the two-sided loop jet and the large filament are reconstructed. Since the two-sided loop jet was along the magnetic field lines of the cavity structure, the trajectory of the two-sided loop jet represents the position of the cavity structure. Based on this assumption, the intersection angle between the cavity's magnetic field and the LF's main axis can be measured by determining the trajectories of the two-sided-loop jet and the main axis of the LF. The paired {\sl SDO} and {\sl STEREO}-A images at 13:10:30 UT are plotted in the top row of \nfig{fig5}, in which the red plus signs and the blue asterisks are the selected positions that used for reconstructing the three-dimensional trajectories of the cavity and the LF, respectively. The reconstructed trajectories of the two-sided-loop jet and the LF are then projected to the east limb, west limb, north polar, and the disk center of the solar disk (see \nfig{fig5} (c -- f)), and they are plotted as red and blue curves, respectively.  We can measure the intersection angle between the cavity and the LF when the trajectories are projected to the disk center. The result reveals that the intersection angle was about 26.7 degrees, well agreement with earlier direct measurement results that the magnetic field of quiescent filament is nearly horizontal, and the magnetic vector is rotated with respect to the filament axis by an angle in the range of 20 to 30 degrees \citep{2003ApJ...598L..67C}.

\section{Conclusion \& Discussion}
In summary, we present the observation of a two-sided loop jet which not only exhibited two distinct reconnection processes but also showed cool erupting material during the formation stage, consistent with the first observation of the MF driven two-sided-loop jet reported by \cite{2019ApJ...871..220S}. Using the high temporal and high spatial resolution stereoscopic imaging observations provided by the {\sl SDO} and {\sl STEREO}-A, for the first time, the two-sided-loop jet was simultaneously observed from two different view angles, and it is evidenced that the two arms of the jet were along the cavity magnetic field lines of a LF. Analysis results of the magnetic fluxes in the eruption source region indicate that the occurrence of the two-sided loop jet was probably triggered by alternative flux emergence and cancellation. The first magnetic reconnection process was due to the interaction between a rising loop and the overlying LF's magnetic field, resembling the physical picture described in the numerical model of two-sided loop jets. The second magnetic reconnection process was due to the rising of a MF located in the eruption source region, which interacted with the overlying LF's magnetic field and triggered the reconnection between them. It is interesting that the trajectory of the two-sided loop was along the magnetic field lines of the cavity structure that hosted the LF at its bottom, which not only traced out the LF's magnetic fields but also injected mass to the LF. Therefore, the present observation provides a new way for understanding the question of filament formation. By taking advantage of the stereoscopic observations taken from two different observation angles, the intersection angle between the magnetic field of the cavity and the main axis of the LF is measured to be about 26.7 degrees. The result not only confirms earlier direct measurement of magnetic field but also provides a new method to diagnose the magnetic structure of solar filaments.

Although early low resolution observations have revealed the coexistence of hot and cool plasma component in some collimated anenome jets \citep{1994ApJ...425..326S,1999SoPh..190..167A,2007A&A...469..331J}, the formation mechanism of the cool component has puzzled solar physicists for many years \citep{1994ApJ...425..326S,1995Natur.375...42Y,2008ApJ...683L..83N}. Until recently, high resolution observations revealed that cool plasma flows are existed in almost all collimated jets \citep{2012ApJ...745..164S, 2015Natur.523..437S}. Due to the explosion nature of the jet base, such kind of solar jets have been renamed to be blowout jets \citep{2010ApJ...720..757M,2012ApJ...745..164S}. Thanks to the recent high resolution observations, it has been evidenced that the cool plasma flow in a blowout jet is actually evolved from the eruption of the MF resided in the eruption source region \citep[e.g.,][]{2012ApJ...745..164S,2017ApJ...851...67S}. To the best of our knowledge, observational evidence on the coexistence of hot and cool components in two-sided loop jets is very scarce, although \cite{1995Natur.375...42Y} have predicted in their numerical model. Therefore, our observation clearly resolved the coexistence of hot and cool components in two-sided loop jets, in which the appearance of the hot component was earlier than the cool one about 3 minutes. In addition, the present event suggests that the formation of two-sided loop jets also involve the eruption of MFs in their source region like collimated blowout jets.

\begin{figure}
\epsscale{1}
\figurenum{6}
\plotone{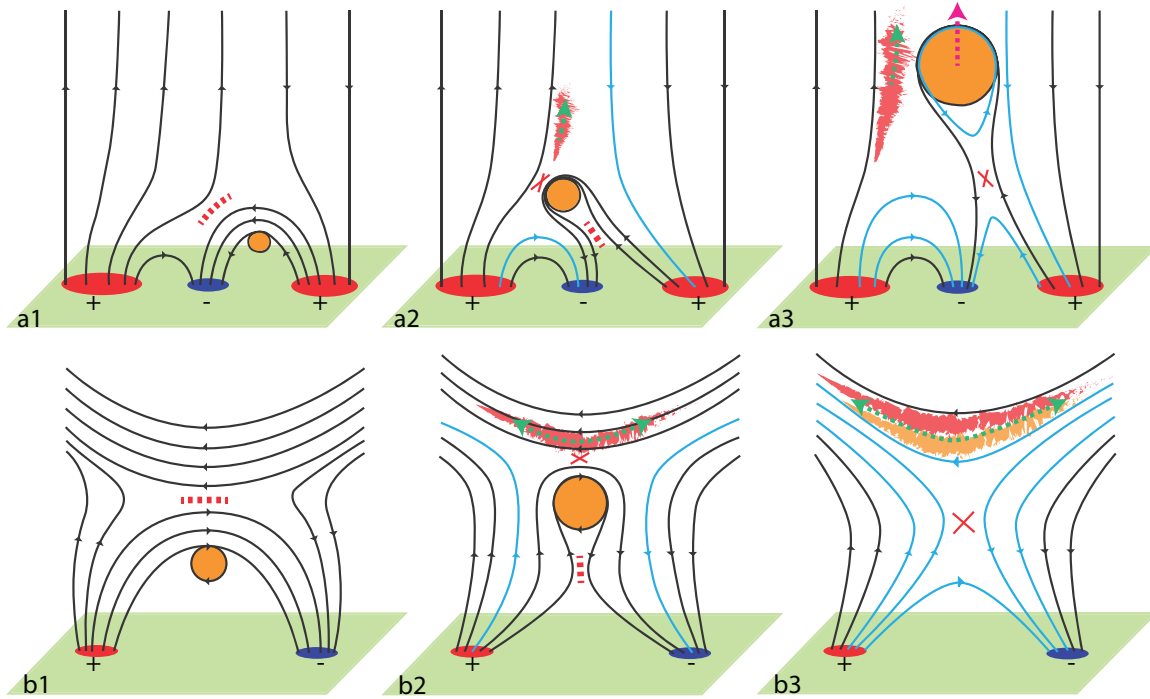}
\caption{Schematic depiction of the eruption processes of collimated blowout jet (top row) and two-sided-loop jets (bottom row). Here, only a few representative field lines are drawn. The red (blue) patches represent positive (negative) polarities, while the lines represent magnetic field lines. On each magnetic field line, the arrow indicates the magnetic field direction. The yellow filled circles represent filament or flux rope. The red dashed curves in a1, a2, b1, and b2 indicate the locations of current sheets, and the red crosses in a2, a3, b2, and b3 indicate  reconnection positions. The cyan lines are newly form field lines after the reconnections. The green arrows in a2 and a3 indicate the jet resulting from the external reconnection, while the red arrow in a3 indicates the erupting MF. The red features in b2 and b3 represent the ejecting hot plasma beams from the first reconnection, and the yellow feature in b3 represents the materials originated from the erupting MF. The green arrows in b2 and b3 indicate the ejecting direction of the two arms of the two-sided loop jet.
\label{fig6}}
\end{figure}

Due to the discovery of blowout jets that contain helical magnetic structures (or MF) in the eruption source region, the explanation model of collimated anenome jets has been upgraded significantly to blowout jet model \citep{1995Natur.375...42Y, 2010ApJ...720..757M}. Especially, the new model of  blowout jets also implies the production of a pair of CMEs as evidenced in some events \citep{2012ApJ...745..164S,2016ApJ...823..129A,2018ApJ...869...39M,2019ApJ...877...61M,2019SoPh..294...68S,2019arXiv190707310D}. For better comparison, we redraw the blowout jet model proposed in \cite{2012ApJ...745..164S} in the top row of \nfig{fig6}. The initial magnetic configuration is consisted of two positive polarities and with a negative polarity in the middle. The basic magnetic field system is composed of two groups of closed loops surrounding by open field lines, and the right closed loop system also contains a MF (see \nfig{fig6}a1). Naturally, a  current sheet exists between the right closed loop and the left open field lines. Due to some instabilities or disturbances such as flux emergence and cancellation, magnetic reconnect will occur in the current sheet. The magnetic energy released by the reconnection will heat the local plasma and produce an outward hot jet along the newly formed field lines. In the meantime, the reconnection also removes the confining field of the MF, which will result in the rising of the MF and the formation of another new current sheet below it  (see \nfig{fig6}a2). The reconnection in this new current sheet will lead to the eruption of the MF at last. In this model, a narrow jet-like CME and a bubble-like CME in the outer corona can be expected, in which the former is developed from the hot jet plasma, while the latter is produced by the erupting filament.

For the formation mechanism of two-sided-loop jets involving the eruption of MF as observed in the present event, we also propose a similar cartoon model as the collimated blowout jets in \nfig{fig6}. The initial magnetic configuration is consisted of a pair of opposite polarities, which are connected by a closed loop system containing a MF. The concave magnetic field of the LF's cavity locates above the closed loop system. Naturally, a current sheet exists between the two magnetic systems (see \nfig{fig6} b1). Due to magnetic activities such as flux emergence and cancellation, the closed loop and the MF will rise and interact with the overlying LF's magnetic field. This will trigger the magnetic reconnection in the current sheet. The magnetic energy release by the reconnection will heat the local plasma and produce the observed small two-sided loop jet in the first eruption stage (see \nfig{fig6} b2). In the meantime, the reconnection also removes the confining magnetic field of the MF, which will result in the fast rising and interaction of the MF with the overlying LF's magnetic field. Therefore, another new current sheet forms underneath the rising MF. During this stage, reconnections occur simultaneously in the current sheets above and underneath the MF. The reconnection above the MF continuing removes the confining field of the MF and release the MF's mass into the concave cavity magnetic field of the LF, while the reconnection occurs underneath the MF produces the flare ribbons at the two opposite polarities that are connected by the newly formed field lines. (see \nfig{fig6} b3). According the present model of two-sided loop jets, the two reconnection processes observed in the present study all occurred between the confining field of the MF and the overlying magnetic field of the LF, they manifested the different reconnection stages in the external current sheet. In the meantime, the observed conjugated flare double ribbons at the two opposite polarities and the eruption of the MF could be evidence of the internal reconnection underneath the MF. It should be pointed out that the basic physical process of our interpretation is similar to the cartoon schematic presented in \cite{2019ApJ...871..220S}, where the rising filament system reconnects with the overlying closed loop system and then the internal reconnection occurs between the two legs of the rising mini-filament. In our case, we propose that the rising filament system first reconnected with the overlying cavity magnetic fields that hosts the LF, then the internal reconnection occurred between the two legs of the stretched confining field lines of the erupting MF. In both case, the two-sided-loop jets are all resulted from MF eruptions and involve two reconnection processes. These results are different to the physical picture described in the classical two-sided-loop jet.

Now, we know that both collimated blowout (or anenome) jets and two-sided loops jets are all involved in the eruption of MFs and two reconnection processes in their formation phases, and jet plasma formed in the two types of jets could be all accelerated by the tension force of the newly formed field lines \citep{1995Natur.375...42Y}. Due to the different magnetic configurations, the two types of jets also show some differences. For a collimated blowout jet, the external reconnection occurs between the MF's confining fields and the ambient open field lines; the jet plasma and the eruption of the MF are all eruptive in nature, and they further respectively develop into a jet-like and a bubble-like CMEs in the outer corona. For a two-sided loop jet, the external reconnection occurs between the confining field of the MF and the overlying horizontal field lines that confines the eruptions of the jet plasma and the MF. Therefore, there would be no CME can be launched in a two-sided loop jet. Furthermore, in relation to the corona heating, the two-sided loop jets should contribute more energy than the collimated blowout jet due to their confining and eruptive nature of the two types of jets.

It has been widely accepted that magnetic field plays a central role in the formation, stability, and eruption of solar filaments \citep{2010SSRv..151..333M}. However, questions on the magnetic structure and material source of solar filaments has not been solved completely  \citep{2015ApJ...803...86Y,2016ApJ...832...23Y,2018SoPh..293...93C}. While observations provide some clues to the magnetic topology, the most direct information of the filament magnetic field comes from the inversion of spectra-polarimetric data \citep{2003A&A...402..769A}. In practice, the direct measuring of the filament magnetic field is very hard. So far, only a few works focus on this topic. Previous direct measurement results indicated that the magnetic field of quiescent (active region) filaments has a strength of 3--15 (30--45) Gauss; the magnetic field in filaments is nearly horizontal to the filaments' main axis, and the magnetic vector is rotated with respect to the filament axis by an angle in the range of 15--30 degrees \citep{1989ASSL..150...77L,2001A&A...375L..39P,2003ASPC..307..468C,2003ApJ...598L..67C}. In the present study, the acute angle between the filament axis and its cavity obtained by the method of three-dimensional reconstruction is about 26.7 degrees, which is not only well consistent with the direct measurement results but also provides a new method for diagnosing the magnetic structure of solar filaments.

The formation of solar filaments include two aspects, namely the formation of the filament channel or magnetic structure, and the formation of filament mass. Previous studies have proposed various models for explaining the mass formation of filaments, such as the injection model \citep{1999ApJ...520L..71W}, levitation model \citep{1994SoPh..155...69R}, and the evaporation-condensation model \citep{1977SoPh...52...37E,2011ApJ...737...27X,2012ApJ...748L..26X,2014ApJ...792L..38X}. \cite{2005ApJ...631L..93L} found that the cool filament mass can be provided the chromospheric jets from one end along the filament main axis (see also \cite{2018ApJ...863..180W}), in which the authors did not explain how the cool jet mass injected into the filament channel. The event studied in the present paper clearly exhibited that the two-sided loop jet injected mass into the overlying LF from the below by the magnetic reconnection between the jet and the magnetic field of the filament's cavity. Therefore, the observation presented here also provides a new way for the understanding of the formation of the filament mass.

\acknowledgments
The authors thank the excellent data provided by the {\sl SDO}, {\sl STEREO}, and GONG teams, and the anonymous referee for comments that improved the paper. This work was supported by the Natural Science Foundation of China (11773068,11633008), the Yunnan Science Foundation (2017FB006), and the West Light Foundation of Chinese Academy of Sciences.

\end{document}